\documentclass[%
 aip, 
 amsmath, amssymb, 
preprint, %
]{revtex4-1}
\usepackage{graphicx}
\usepackage{dcolumn}
\usepackage{bm}
\DeclareUnicodeCharacter{2212}{-}
\usepackage[utf8]{inputenc}
\usepackage[T1]{fontenc}
\usepackage{mathptmx}
\usepackage{etoolbox}
\usepackage{wrapfig}
\usepackage{amssymb,amsmath}
\usepackage{enumitem}
\usepackage{xcolor}
\usepackage{natbib}
\usepackage[bottom]{footmisc}
\usepackage{hyperref}
\usepackage[utf8]{inputenc}

\usepackage{ltablex}
\usepackage{siunitx}
\usepackage[flushleft]{threeparttablex}
\usepackage{listings}

\makeatletter
\def\@email#1#2{%
 \endgroup
 \patchcmd{\titleblock@produce}
  {\frontmatter@RRAPformat}
  {\frontmatter@RRAPformat{\produce@RRAP{*#1\href{mailto:#2}{#2}}}\frontmatter@RRAPformat}
  {}{}
}%
\makeatother

\definecolor{mygreen}{rgb}{0,0.6,0}
\definecolor{mygray}{rgb}{0.47,0.47,0.33}
\definecolor{myorange}{rgb}{0.8,0.4,0}
\definecolor{mywhite}{rgb}{0.98,0.98,0.98}
\definecolor{myblue}{rgb}{0.01,0.61,0.98}
\begin{document}
\title{Force and torque-free helical tail robot to study low Reynolds number microorganism swimming}

\author{Asimanshu Das}
\affiliation{Center for Fluid Mechanics,  Brown University,  Providence,  RI 02912,  USA}
\author{Matthew Styslinger}
\affiliation{Center for Fluid Mechanics,  Brown University,  Providence,  RI 02912,  USA}
\author{Daniel M Harris}
\affiliation{Center for Fluid Mechanics,  Brown University,  Providence,  RI 02912,  USA}
\author{Roberto Zenit}\thanks{roberto\_zenit@brown.edu}
\affiliation{Center for Fluid Mechanics,  Brown University,  Providence,  RI 02912,  USA}

\date{\today}

\begin{abstract}

Helical propulsion is used by many microorganisms to swim in viscous-dominated environments. Their swimming dynamics are relatively well understood,  but detailed study of the flow fields and actuation mechanisms are still needed to realize wall effects and hydrodynamic interactions. In this letter,  we describe the development of an autonomous swimming robot with a helical tail that operates in the Stokes regime. The device uses a battery-based power system with a miniature motor that imposes a rotational speed to a helical tail. The speed,  direction,  and activation are controlled electronically using an infrared remote control. Since the robot is about 5 centimeters long,  we use highly viscous fluids to match the Reynolds number to be $\text{Re} \lessapprox 0.1$.  Measurements of swimming speeds are conducted for a range of helical wavelengths,  $\lambda$, head geometries and rotation rates,  $\omega$. We provide comparisons of the experimental measurements with analytical predictions derived from resistive force theory. This force and torque-free neutrally-buoyant swimmer mimics the swimming strategy of bacteria more closely than previously used designs and offers a lot of potential for future applications.
\end{abstract}

\maketitle

Microorganism propulsion mechanisms show a fascinating variability which reflect in their evolutionary adaptations \cite{ramsay2011quorum, foster1984rhodopsin, chicurel2000slimebusters}. Usually,  microorganisms such as bacteria are a few microns in size and dwell in aqueous fluids. At such scales,  their swimming Reynolds number typically ranges from $10^{-4}$ to $10^{-1}$; hence,  the dynamics of swimming is different from that at a macroscopic scale\cite{lauga2009hydrodynamics, hancock1953self}. Time-reversibility prevails in such viscous dominated flows,  wherein reciprocal motion cannot result in net propulsion. To circumvent this issue,  two common strategies used by microorganisms are  i) rotation of helical flagella analogous to a corkscrew motion; ii) undulation of the flagella resembling a beating motion\cite{purcell1977life}. The majority of motile bacteria rotate helical flagellar filaments in order to achieve locomotion. Escherichia coli (E. coli) use a helical bundle of flagella of a few micrometers in length and 20 nm in diameter which they rotate at a frequency of 100 Hz\cite{gray1955propulsion}. Using living cells to conduct detailed measurements to unveil the mechanisms of propulsion is extremely challenging. In addition to their small size which leads to visualization challenges,  it is difficult to isolate particular parameters that affect swimming performance. Hence,  one must rely on either computer simulations, theoretical calculations, or synthetic devices that mimic the motion of such organisms\cite{hyon2012wiggling,  arco2014viscous}. For the case of biomimetic robots,  the challenge is to use devices that retain sufficient details of the swimming strategy. Most devices used to date rely on an external power source,  such as external magnetic field actuation\cite{espinosa2013fluid,gomez2017helical, angeles2021front}. Although swimming is achieved and changes in the shape of the helical tail can be controlled and varied,  there are several important limiting factors. These devices are actuated by an external field  and are not torque-free; hence,  they cannot be considered self-propelled. An important consequence of this condition is that the head and the tail rotate in the same direction in these devices. Therefore,  the flow produced around them is most likely quite different from that in real bacteria. Consequently, the interaction between swimmers and walls is also expected to be significantly different. 


In this paper,  we build a remote-controlled macroscopic torque and force-free swimmer to remedy some of the limitations described above. The robot presented here is closer to the natural system,  allowing flexibility to vary the head and tail geometries. We can thus directly assess the importance of geometric and kinematic factors on the swimmer's performance. In particular, we measure swimming speeds for different rotation rates and, head and tail geometries. We compare these measurements to predictions of an analytical model based on resistive force theory and find a good agreement. Note that recently Kroo et al. \cite{kroo2021swimming} used a swimmer with similar characteristics to that described here. Their focus was on the so-called snowman swimmer\cite{Pak2012,Puente2019} and its application as a rheometer.

\begin{figure}[!htp]
	\centering{
		\includegraphics[width=0.8\textwidth]{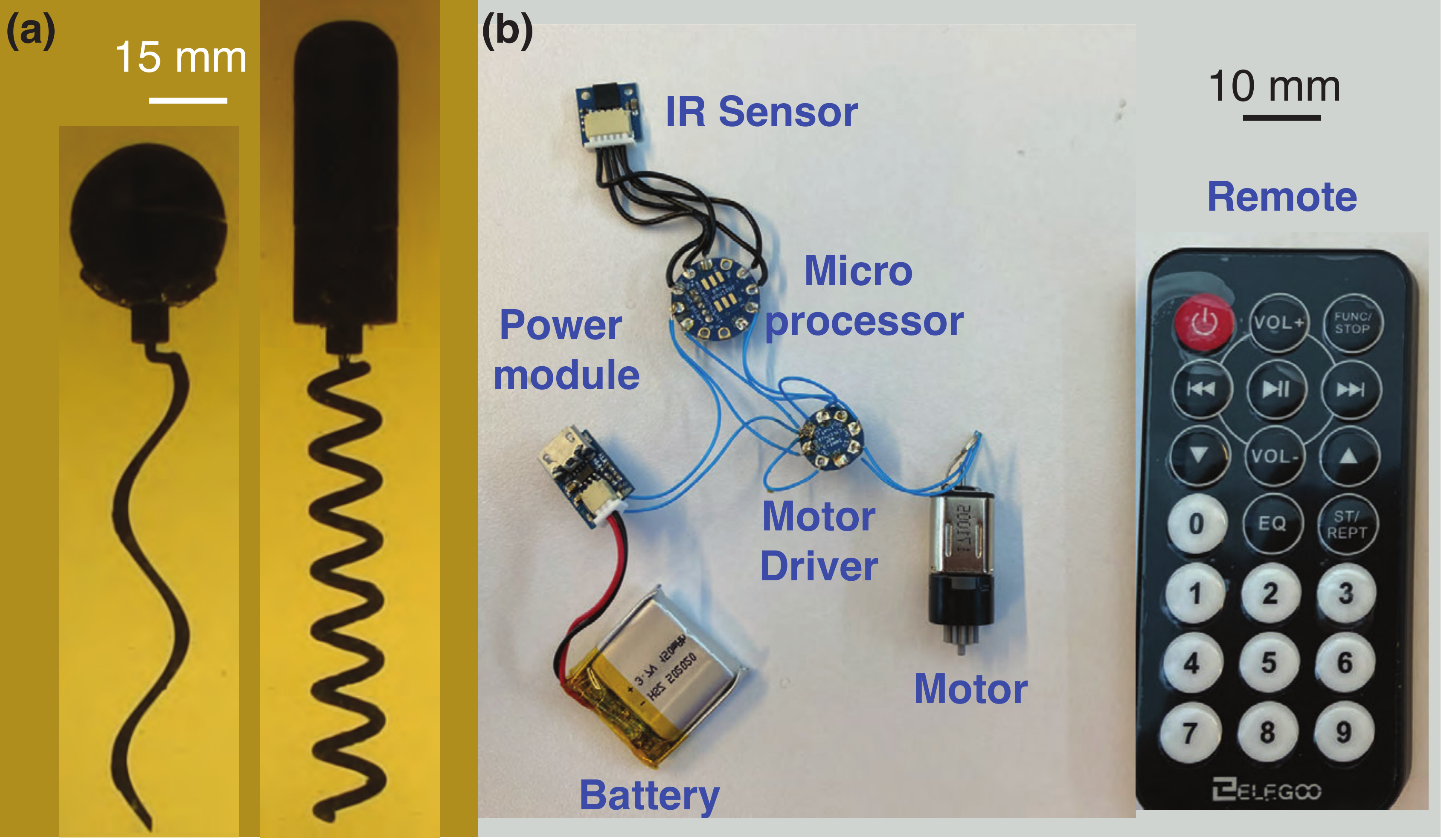}}
	\caption{ (a) Snapshots of helical swimmers with spherical and cylindrical heads. Scale bar gives an estimate of the size of the swimmers. (b) The electronic circuit used to operate the helical swimmer with its various components soldered together. }
	\label{fig1:CAD + electronics}
\end{figure}


The design consists of two subsystems - mechanical and electrical. The swimmer aims to emulate bacteria swimming,  so it consists of a head and a rotating helical tail.   Fig. \ref{fig1:CAD + electronics}(a) shows the mechanical design of the swimmer with helical tails. In order to mimic a bacterium, 
the robotic swimmer's head and tail are geometrically similar and has scaled-down dimensions of an E. coli. The head is cylindrical with a rounded front end, 20 mm in diameter and 60 mm long. Additionally, tests were conducted using a spherical head geometry, of 30 mm in diameter. The head was 3-D printed (Formlabs Form 2 printer, Clear resin, 25 micron print resolution), designed as a hollow shell assembled from two sealable parts. The electronics and motor were fully embedded within the head. The motor shaft shield protrudes from the head, such that the tail could be connected directly to the teethed shaft of the motor. The tail, which was also 3-D printed, was tightly fitted to minimize wobbling motion due to misalignment. The tail is 90 mm long, with a filament diameter of 2.5 mm and is 12 mm in diameter with . The CAD models of all components can be found in the supplementary material.
A high-torque geared motor actuated the tail.
In addition to the mechanical design,  the electronics play a crucial role in the present device. We used a miniature open-source circuit (from Tiny Circuits) that fit into our millimeter scale swimmer. The electronic circuit consisted of a power source (70 mAh Li-Po battery),  a motor driver circuit,  a motor controller circuit,  an IR sensor,  and the motor. The power module and battery supplies 3.7 V while the microprocessor integrated circuit ensures signal processing and logic control. The motor driver chip acts as a bridge between the processor and the motor and regulates its speed and rotation direction. The microprocessor was programmed to control the motor's speed based on the voltage it draws from the power source. Fig. \ref{fig1:CAD + electronics}(b) illustrates the electronic circuit (wiring connections) used and the remote control used to operate the swimmer. The various buttons in the remote were programmed to switch the speed and rotation direction.

 \begin{figure}[!htp]
	\centering{
		\includegraphics[width=0.7\textwidth]{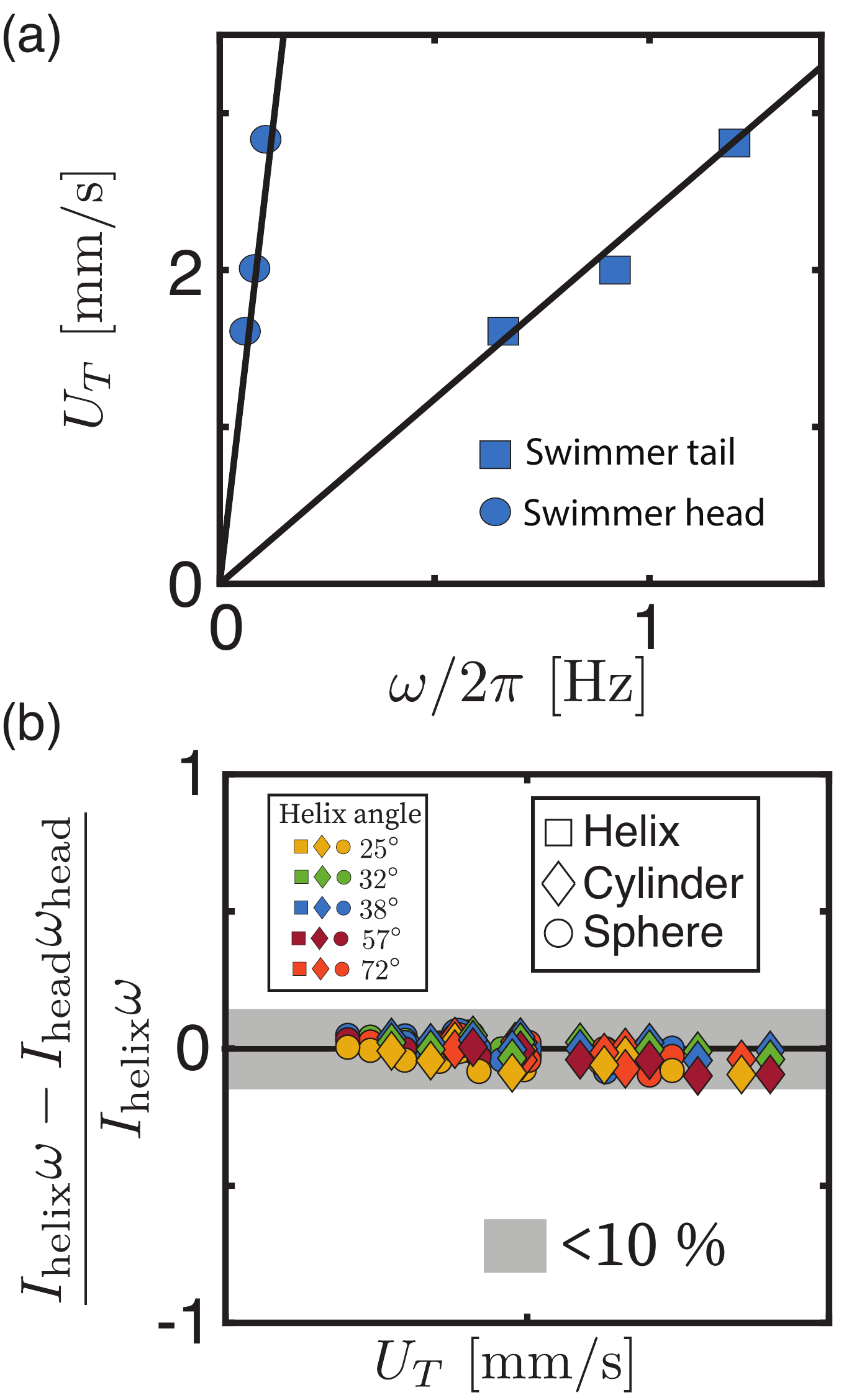}}
	\caption{ (a) Swimming speed vs. frequency of rotation for a spherical head swimmer with a helical tail with $\theta=57 ^\circ$ . The expected linear relationship between the quantities is confirmed by the linear fits,  which has been earlier established for a swimmer swimming in a Newtonian fluid \cite{lauga2009hydrodynamics}. (b) The net normalised angular momentum with swimming speed. The error bars are smaller than symbol size.}
	\label{fig2:Torque-free}
\end{figure}
\begin{figure}[!htp]
	\centering{
		\includegraphics[width=0.7\textwidth]{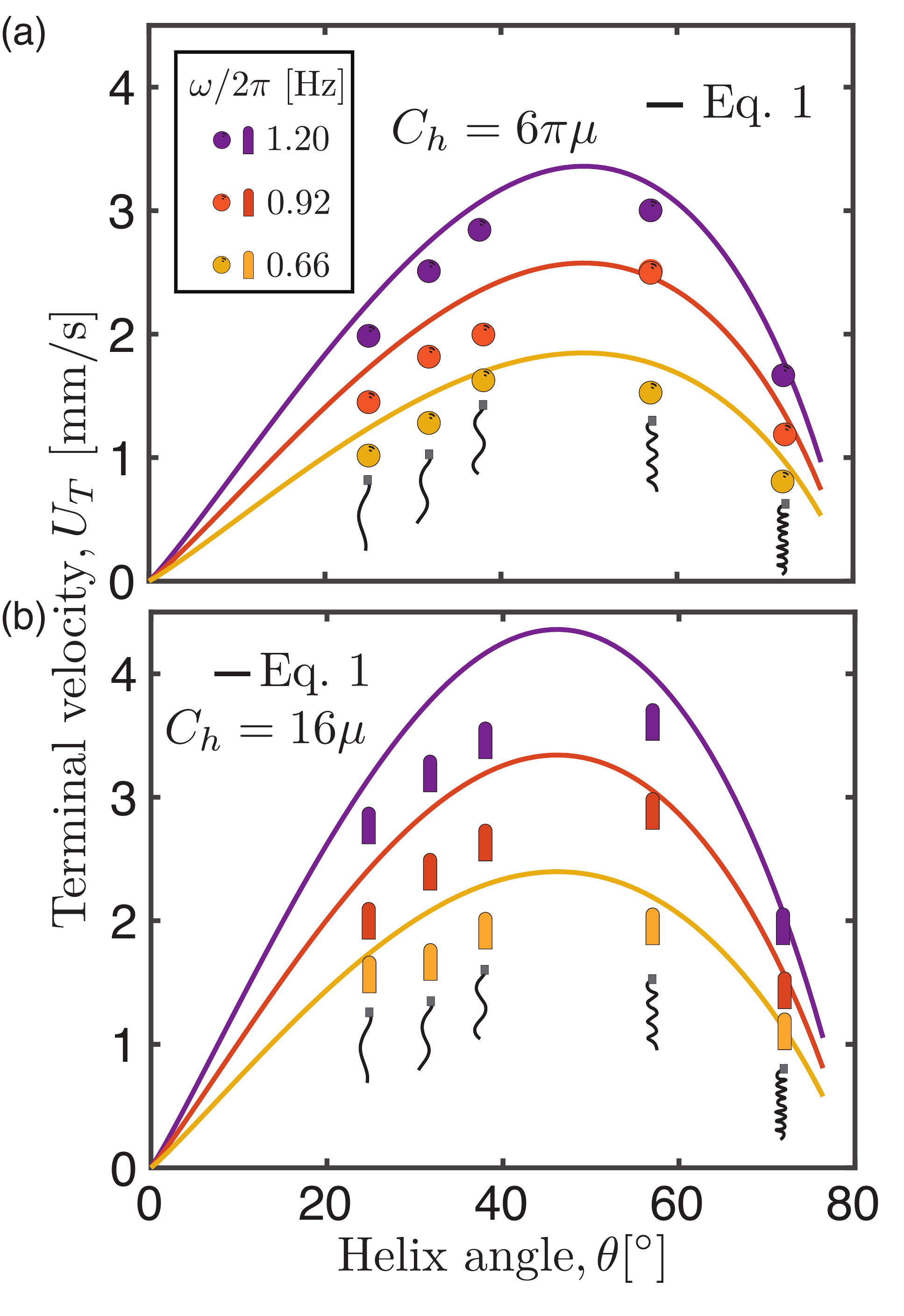}}
	\caption{ (a) The swimming speed,  $U_T$,  with various pitch angles,  $\theta$ for swimmers for (a) spherical and (b) cylindrical head geometry. The tail length, $L$, is 90 mm and its helix diameter, $d$, is 12 mm across all helices. Circles represent the swimmer with spherical head and the elongated symbols represent data from the cylindrical head swimmers. The drawing of helices gives a representation of the corresponding helix profiles. The solid lines gives model predictions from resistive force theory. Note: The error bars are smaller than the symbol size in both (a) and (b). }
	\label{fig1:Model+expt}
\end{figure}

Since the robot is fully immersed in a fluid, there were a few additional protocols taken to prevent any leakage into the robot and to ensure hydrostatic stability along-with neutral buoyancy. The details of these procedure are  mentioned in the supplementary material \cite{supp}. The robot was immersed in a high-viscosity corn syrup water solution ($\mu$, $1 \text{Pa.s}$)  to match a low Reynolds number condition which ensures geometric,  kinematic,  and dynamic similarity with  real bacteria. The Reynolds number is defined based on the swimming speed and head diameter which is defined as $\frac{\rho U D}{\mu}$ which had a maximum value of 0.1 for all the measurements shown here. The experiments were conducted with two different head shapes,  several helical tails with pitch angles,  $\theta$,  ranging between $25^{\circ}$ and $72^{\circ}$ (corresponding to  helix wavelengths from 6 to 72 mm) and for a range of rotational speeds, from 0.3 to 1.2 Hz. The swimmer's motion was recorded using a digital camera at 120 fps (Nikon D5600,  24.2 megapixel resolution),  illuminated with a uniform back-light. The swimmer was painted with matte black paint to ensure good contrast. A mesh grid pattern,  with a grid size of 10 mm,  was used for calibration. The {\it{Canny}} edge detection algorithm was employed in Matlab to extract the head profile and its centroid allowed to accurately measure the velocity. We put a few markers on the swimmer's head and tail to track the angular speed. A sample video of the experiments is included in the supplemental materials\cite{supp}. Figure \ref{fig2:Torque-free}(a) shows  the swimming speed,  $U_T$,  as a function of the angular rotation velocity,  $\omega$,  of the head (squares) and tail(circles). The fit lines shows a linear dependence of both the quantities for a Newtonian fluid which is well known \cite{lauga2009hydrodynamics}. In order to verify the torque-free nature of our swimmer we can consider the conservation of angular momentum for the head and the tail: the calculated the moment of inertia and the measured angular velocity are balanced. Here we see that $I_{head}\omega_{head}-I_{tail}\omega_{tail}$ $\approx$ 0,  which ignores the added rotational inertia of the surrounding fluid since Re is small. Fig. \ref{fig2:Torque-free}(b) shows the difference of angular momentum for different head and tail geometries. The low values shown in the figure ($<10\%$) indicate the torque-free nature of the robot.

To further validate the swimming performance of the present device,  we compare the swimming speeds with predictions of a model based on resistive force theory. Assuming neutral buoyancy of the swimmer,  the helix rotation produces a thrust that propels
the swimmer; the thrust is balanced by the drag experienced by both tail and head.  The swimming velocity for the swimmer can be found analytically\cite{lighthill1976flagellar, rodenborn2013propulsion} and is given by the expression 
\begin{equation}
    U_T=\frac{\omega R(C_n-C_t)\sin \theta \cos \theta }{(C_n\sin^2 
\theta+ C_t \cos^2 \theta) (L/\cos \theta)+ C_{h} R_h }{(L/\cos \theta)}, 
\label{eqn: Drag spherical head}
\end{equation}
where $C_n$ and $C_t$ are the Gray and Hancock  normal and tangential drag coefficients \cite{gray1955propulsion}, respectively,  considering a large cell body. Here $L$ is the length of the helix,  $\theta$ is the pitch angle,  $R$ is the helix diameter; $R_h$ and $C_h$ are the radius and drag coefficient of the head, respectively. Fig. \ref{fig1:Model+expt}(a) and (b) show the comparison between experimental measurements with  predictions from Eq.\ref{eqn: Drag spherical head},  considering  spherical and cylindrical heads, respectively. For the case of a sphere, the viscous drag is known exactly\cite{maxworthy1965accurate}, $C_h=6 \pi \mu$; and the model and measurements show good agreement. For the cylindrical head the drag coefficient is not known exactly for the head shape that we used. Using an empirical relation based on the aspect ratio\cite{huner1977cylinder} we estimate  $C_h\approx 16\mu $. In spite of this approximation,  the comparison is very good. In both cases as we see in Fig.~\ref{fig1:Model+expt}~(a) and (b), the swimming speed is slightly overpredicted which might be due to the interactions between the counter rotating head and tail creating a complex flow near the connection of the cell body and flagella, and it has not been accounted for in the model. 

In summary,  we have developed a remote-controlled torque and force-free robotic swimmer which is capable of emulating bacterial swimming. We quantified the swimmer's propulsion for different head shapes and  helix profiles over a range of swimming speeds. A simple model based on the resistive force theory was developed,  which captured the dynamics of the swimmer well for the case of a Newtonian fluid. Most importantly,  the swimmer's compact design and versatility can be used for a variety of applications. In particular,  the device can be utilized to study bacterial motion where the details of the velocity field are essential such as collective swimming and interaction with walls and interfaces. Furthermore,  the current design can easily be modified to consider several flagella,  flexible flagella,  and flexible head-tail couplings\cite{kim2004particle,  qu2018changes,  tournus2015flexibility}. Exploring such effects can significantly extend our current understanding of bacterial swimming dynamics.

\section*{Supplementary Material}
The supplementary material of this note has the bill of materials, all of the necessary CAD files, detailed construction instructions and code for electronic control. It also contains details of the theoretical model.

\section*{Acknowledgements}
We acknowledge Varghese Mathai for useful discussions. We also thank Ben Lyons and George Worth for assistance with the overall development of the robot.
\section*{Data Availability}
The data that supports the findings of this study are available within the article and its 
supplementary material. 

\bibliographystyle{apsrev4-1}
\bibliography{reference}
\clearpage
\newpage

\appendix

\clearpage
\newpage

\section*{Supplemental Material}

\section{Bill of materials}
The list of parts used to build the device can be found on Table \ref{table:materials}.

    \begin{table*}[b]
    \centering
\caption{Bill of materials. Hyperlinks with are provided for each item.}
    \label{tab:freq}
\caption*{\footnotesize
      
            }
  \begin{tabular*}{\linewidth}{S @{\extracolsep{\fill}}
                            *{4}{S[group-separator = {,},
                                   group-minimum-digits = 4
                                    ]}}\\
      \toprule
  {Sl No.} & {Part Name} & {Description} \\
  1       & \color{blue}{\href{https://tinycircuits.com/collections/accessories/products/micro-usb-cable-3-feet}{Micro USB cable}  } & {For connecting the computer to mini-processor} \\
  2       & \color{blue}{\href{https://tinycircuits.com/collections/all/products/tinylily-motor-board}{TinyLily motor board}  } & {Motor controller circuit board}\\
  3      &\color{blue}{ \href{https://tinycircuits.com/collections/batteries/products/lithium-ion-polymer-battery-3-7v-70mah}{Lithium ion polymer battery - 3.7V 70MAH}  } & {Powers the circuit} \\
  4       & \color{blue}{\href{https://tinycircuits.com/collections/tinylily/products/tiny-battery-charger}{Tiny battery charger}  } & {To recharge the battery} \\
  5      &\color{blue}{ \href{https://tinycircuits.com/collections/tinylily/products/tinylily-mini-processor}{TinyLily mini processor}  } & {It is the processing unit of the circuit} \\
  6      & \color{blue}{\href{https://tinycircuits.com/collections/wirelings/products/ir-receiver-38khz-wireling}{IR receiver wireling}  }  & {IR sensor} \\
  7     & \color{blue}{\href{https://www.amazon.com/dp/B07N7PY71F?ref=ppx_pop_dt_b_product_details&th=1}{Micro gearbox motor}  } & {Motor}  \\
  8     & \color{blue}{\href{https://www.adafruit.com/product/1970}{Connecting wires}  }  & {Wires for soldering connections}  \\
  9     & \color{blue}{\href{https://www.acehardware.com/departments/automotive-rv-and-marine/fluids-and-lubrication/automotive-lubrication-greases/87211?store=06234&utm_source=google&utm_medium=cpc&gclid=CjwKCAiA1uKMBhAGEiwAxzvX9-bqrL3BjYtoTr9fLIlxfdURV1KOExQw9dx7OHkTZYsyHfsBivayYRoCA00QAvD_BwE&gclsrc=aw.ds}{AGS grease} }   & {Grease to prevent leak}  \\
  10    & \color{blue}{\href{https://tinycircuits.com/collections/wireling-accessories/products/5-pin-extension-cable}{Wireling connections}}  & {Connecting unit for the IR sensor}   \\
   11    & \color{blue}{\href{https://www.amazon.com/Dixon-Valve-TTB75-Industrial-Temperature/dp/B003D7K8E0/ref=sr_1_5?keywords=plumbing+tape&qid=1637486598&sr=8-5}{Sealant tape}}  & {Kapton tape for sealing the head} \\
   12    & \color{blue}{\href{https://formlabs.com/store/clear-resin/}{Formlabs Clear Resin}}  & {Resin for 3D printing}   \\
  
  \end{tabular*}
  \label{table:materials}
\end{table*}

\section{CAD files} 

The CAD files for the swimmer head and tail can be found here: \href{https://www.dropbox.com/sh/6g6v7bekx6t7fob/AAB5JDrt20Sc247j_7Hyk1--a?dl=0}{\tt [link]}

\section{Supplemental Videos}

Supplemental video S1: Cylindrical head swimmer with helical tail angle, $\theta=72^\circ$. \href{https://www.dropbox.com/s/ogsud0jzyuyetc0/Cylindrical_head.mov?dl=0}{\tt[link]}

Supplemental video S2: Spherical head swimmer with helical tail angle, $\theta=38^\circ$. \href{https://www.dropbox.com/s/25qf3e2a29604b4/Spherical_head.mov?dl=0}{\tt[link]}

Supplemental video S3: The stability of the swimmer is tested with external perturbations. \href{https://www.dropbox.com/s/y0q8s1ejfd8njdg/Stabilty_Video.mov?dl=0}{\tt[link]}

\section{Leakproofing and electronics care}

As the experiments are carried out by submerging the robot in a viscous fluid, it is important to ensure the device is leakproof.   We used a thin strip of Kaptan tape, on the threaded joint of the head for this purpose. There are more robust ways of the sealing it like silicone, epoxy or any leak proofing clay but i) it creates a uneven surface if not properly coated and ii) it makes it more permanent and it is difficult to open the swimmer and access the electronics. Another opening is where the motor is attached to the tail, the gap between shaft shield and the casing possibly can damage and make it malfunction. In order to prevent that we apply a thick viscous grease (AGS white lithium grease) in that joint, which prevents the other fluid from seeping in.

\section{Swimmer Stability}

The concept of hydrostatic stability is deemed as one of the most crucial area of focus in designing a underwater swimmer, a ship or anything that needs to be in fully or partially immersed in a fluid. In these set of experiments we needed a neutrally buoyant swimmer, that swims relatively far from the tank walls and the fluid surface. Along with achieving neutral buoyancy by balancing the buoyant force and the weight, we needed to ensure that the swimmer is in a state of stable equilibrium. It is well known that the swimmer can show some wobbling if the helical axis is not perfectly aligned with the head. We needed to ensure the swimmer is capable of maintaining its stability irrespective of any flow-induced perturbations and small asymmetries in the manufactured design. 

\subsection{Neutral buoyancy}

One of the important consideration in designing the swimmer head was to fully accommodate the electronic circuit. Initially the swimmer naturally buoyant and floats, which posed as a challenge for us to add weights later. In order to achieve this we had to make subtle changes in the swimmer's head design. We added a compartment in the swimmer's head to fit weights of some kind. We used standard brass nuts, spacers and pins to add significant amount of weight, and resorted to using iron filings for more fine adjustments. We glued the bigger weights in the swimmer head to make it reproducible, as well as to make sure it doesn't reposition in the compartment and affect the swimmer dynamics. The iron filings were sprinkled inside the head and it had no effect on the swimming speed measurements. In practice, it is very difficult to achieve perfect neutral buoyancy but we ensure the settling/rising speed is at least ten times smaller than the lowest propulsion speed measured. This was one of the important challenges in the experiment, and the real bacteria can exchange gases with the surroundings to control its depth.In order to allow for the gravity center to pass through the buoyant center we orient to vertical swimming which balances the torque. But if the robot is used in a horizontal setting additional consideration to balance that torque which is the result of the difference in the axis of the buoyant force and weight. All these calculations of the position of center of mass and center of buoyancy can be made from the design, and it can give us an accurate estimate of the additional weight needed.

\subsection{State of stable equilibrium in vertical swimming}
As mentioned above, in order ensure stabilty from perturbations in the flow or rising from small asymmetries we needed to achieve a stable equilibrium. If the center of mass is below the center of buoyancy, the system restores itself from any small disturbance. In order to achieve this we added most weights to the lower compartment of the head. As it is fully submerged it is easy to estimate the center of buoyancy and it remains the same independent of the additional mass. The center of mass can be calculated from the mass distribution, or experimentally found by balancing the body. We also share a video for stability check. The experiments performed here are conducted in a vertical setting; in this manner, the buoyant and mass centers were located along the same line.  As our robot swims bottom to top, the weights were added to the lower part of the swimmer's head such that the center of mass is below the center of buoyancy to make the swimmer stable. This allowed the swimmer to realign vertically,  in case of small perturbations while swimming.

\begin{figure}[!htp]
	\centering{
		\includegraphics[width=0.5\textwidth]{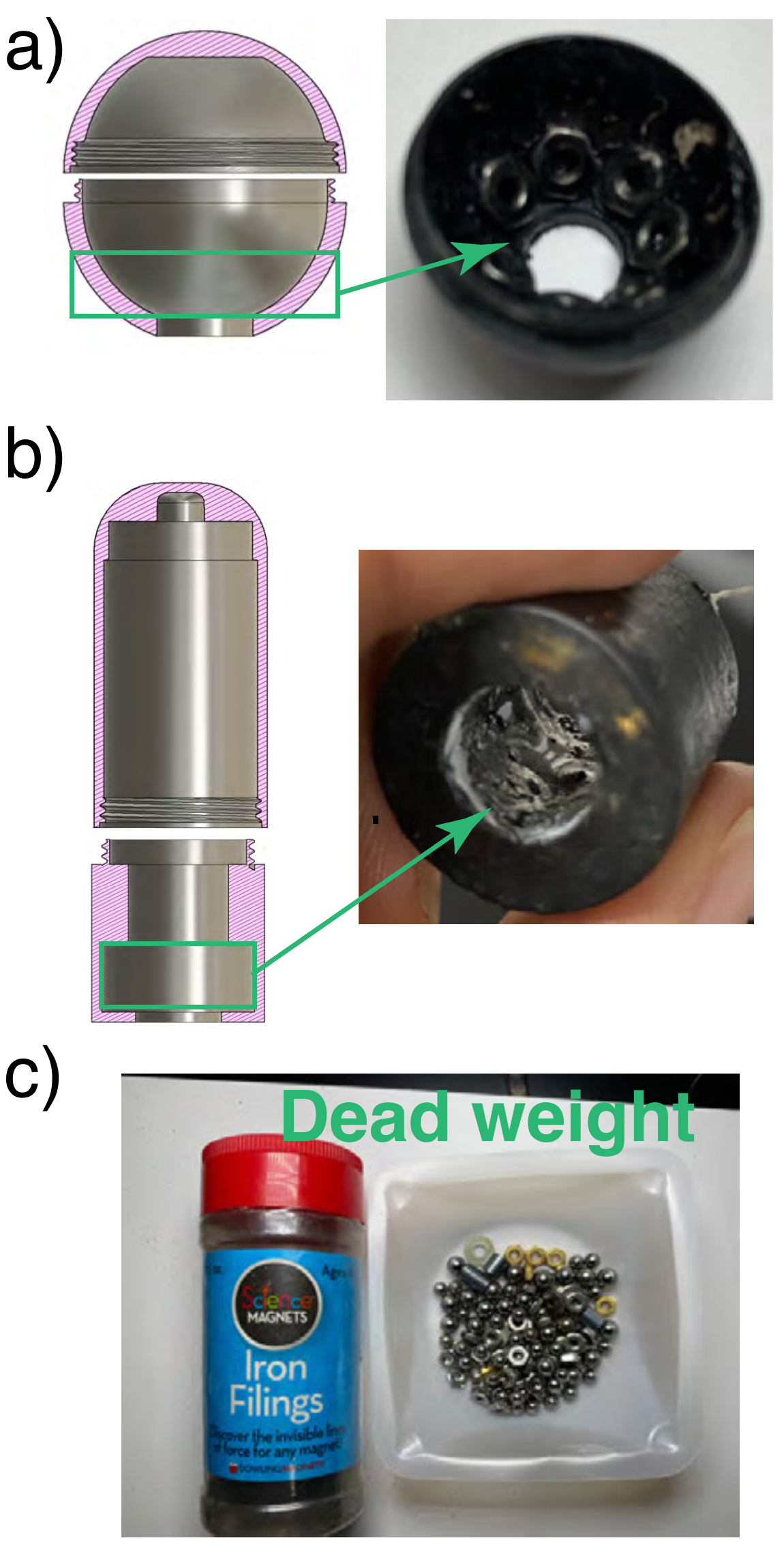}}
	\caption{(a) A section view of the spherical head design. Added weights in the bottom part of the head.(b) A section view of the cylindrical head design. Added weights in the bottom part of the head.(a) Representative dead weights and iron filings for finer balance.}
	\label{figS:Dead weights}
\end{figure}

\section{Theoretical modeling}
The translation of any rigid slender body through a viscous fluid can be analyzed, with an assumption that the radius of curvature of the body is large compared with the body radius. The theory allows us to obtain the total force and torque for motion of a flagellum by integrating the local forces on each small segment. The helix is the body of interest which in our case has a much larger radius of curvature compared to the filament radius. We consider the drag coefficients in the normal and tangential directions, $C_n$ and $C_t$ respectively in estimating the local forces, and the drag on the head:
$$\text{Thrust},~ F= (\Omega R)(C_n-C_t)\sin \theta \cos \theta \frac{L}{\cos \theta}$$
$$\text{Drag},~ D_{\text{helix}}= U(C_n \sin^2 \theta + C_t \cos^2 \theta)\frac{L}{\cos \theta}$$
$$ \text{Head drag},~ D_{\text{head}}= U C_h R_h.$$

The total drag on the swimmer is estimated considering $D=D_{\text{head}}+D_{\text{helix}}$
The sets of drag coefficients are commonly used in the literature those by Gray and Hancock \cite{hancock1953self} and those by Lighthill\cite{lighthill1976flagellar}; are based on slender body theory. Comparing both the coefficients, we find that Gray and Hancock takes into consideration the effect of a large cell body, and provides more conservative estimates. The Lighthill predictions are more suitable for flagellum without a cell body. In this set of experiments we have a significant cell body due to which we use the Gray and Hancock coefficients,
$$C_t=\frac{2\pi \mu}{\text{ln}\frac{2\lambda}{a}-1/2}, ~~~C_n=\frac{4\pi \mu}{\text{ln}\frac{2\lambda}{a}+1/2} $$
where $\lambda$ is the wave-length of the helix, $a$ is the filament radius, and $\mu$ is the fluid viscosity.

For the head, solutions can be readily found for simple shapes. For a sphere, we have:
$$C_{{h}}=6\pi \mu.$$

As the helical tail rotates, the swimmer starts from rest and accelerates until the counter acting drag force suppresses it and it reaches a terminal velocity. In order to model the dynamics,  we can write, $F-D=(m+m_a) \frac{dU}{dt}$, where $m$ is the mass of the body and $m_a$ is the added mass, and $dU/dt$ is the acceleration . 

Therefore, 
\begin{equation}
    F-A_{\text{head}}U=(m+m_a)\frac{dU}{dt}
\end{equation}
where $A_{\text{head-spherical}}=(C_n \sin^2 \theta + C_t \cos^2 \theta)({{L}/{\cos \theta}})+ C_{h} R$. For a spherical head $C_{h}=6\pi\mu $, while for a cylindrical head (considering the dimensions of our robot) $C_{h}=16 \mu $. 

Solving the above differential equation, with an initial condition that the swimmer starts from rest i.e $U(0)=0$, we obtain
\begin{equation}
  U=\frac{F}{A_{\text{head}}} \left(1-e^{-\frac{A_{\text{head}}}{(m+m_a)}t}\right)  
\end{equation}
To find the terminal velocity, we can take the limit as $t\rightarrow \infty$, of the expression above to obtain that $U_T=\frac{F}{A_{\text{head}}}$. Note also that for the conditions considered here, the ratio ${A_{\text{head}}}/{(m+m_a)}$ is very large. For this case, the swimmers attains its terminal speed almost instantly in about 0.2 seconds. This is expected, as the flow is dominated by viscous effects (small Re).

\section{Arduino code for motor control}

\begin{lstlisting}

#include <IRremote.h>

#include <Wire.h>
#include <Wireling.h>

#if defined (ARDUINO_ARCH_AVR)
#define SerialMonitorInterface Serial
#elif defined(ARDUINO_ARCH_SAMD)
#define SerialMonitorInterface SerialUSB
#endif

// Define codes as 32 bit unsigned integers
uint32_t powerCode = 0xFFA25D;
uint32_t func_stop = 0xFFE21D;
uint32_t first_button = 0xFF30CF;
uint32_t second_button = 0xFF18E7;
uint32_t third_button = 0xFF7A85;
uint32_t fourth_button=0xFF10EF;
uint32_t fifth_button=0xFF38C7;
uint32_t sixth_button=0xFF5AA5;
uint32_t seventh_button=0xFF42BD;
uint32_t muteCode = 0xFF7A86;


int motorDirPin = 2;      // Motor direction connected to digital pin 2
int motorSpeedPin = 3;    // Motor speed connected to analog pin 3

// Receive and transmit can be done on any IO pin. Pick A0-A3 for Wireling ports 0-3.
int RECV_PIN = A1;

IRrecv irrecv(RECV_PIN);

void setup(void) {
  SerialMonitorInterface.begin(9600);
  Wire.begin();
  Wireling.begin();
   while (!SerialMonitorInterface && millis() < 5000); //This will block for 5 seconds or until the Serial Monitor is opened on TinyScreen+/TinyZero platform!

  irrecv.enableIRIn(); // Start receiving data

  pinMode(motorDirPin, OUTPUT);       // sets the pin as output
  pinMode(motorSpeedPin, OUTPUT);     // sets the pin as output
digitalWrite(motorDirPin, LOW);
digitalWrite(motorSpeedPin, LOW);
}


void loop() {
  decode_results results;
  if (irrecv.decode(&results)) {
    irrecv.resume(); // Receive the next value
    if (results.decode_type = NEC && results.bits == 32) { //Check if there's a match for our expected protocol and bitcount 
      if (results.value == powerCode) {
        SerialMonitorInterface.println("It's running");
                {
                   delay(50);
                   digitalWrite(motorDirPin, LOW);  
                   delay(50)    ;
                  analogWrite(motorSpeedPin,255); 
                  delay(50);// Max speed forward 
                }
      } else if (results.value == func_stop) {
        SerialMonitorInterface.println("Stop running");
        
         {  
                   delay(50); 
                   digitalWrite(motorDirPin, LOW);  
                   delay(50)  ;
                  analogWrite(motorSpeedPin,0);
                  delay(50);   
                }
        
      } else if (results.value == first_button) {
        SerialMonitorInterface.println("Speed_1"); 
                {
                 
                    delay(50); 
                   digitalWrite(motorDirPin, LOW);  
                   delay(50)  ;
     analogWrite(motorSpeedPin,100);
     delay(50);
                  
                }
        
      } else if (results.value == second_button) {
        SerialMonitorInterface.println("Speed_2");
                    {
                      // Ramp the motor speed up
                      delay(50);
                      digitalWrite(motorDirPin, LOW);  
                   delay(50) ; 
analogWrite(motorSpeedPin,150);
     delay(50);
 
                    }
      } else if (results.value == third_button) {
        delay(50); 
        digitalWrite(motorDirPin, LOW);  
                   delay(50)  ;
        SerialMonitorInterface.println("Speed_3");
               analogWrite(motorSpeedPin,180);
     delay(50);
      }
      else if (results.value == fourth_button) {
        delay(50); 
        digitalWrite(motorDirPin, HIGH);
        delay(50);
        analogWrite(motorSpeedPin,155); 
        SerialMonitorInterface.println("Speed_4_opposite_direction");
     delay(50);
      }  

      else if (results.value == fifth_button) {
        delay(50); 
        digitalWrite(motorDirPin, HIGH);
        analogWrite(motorSpeedPin,105); 
        SerialMonitorInterface.println("Speed_5_opposite_direction");
     delay(50);
      }  
      else if (results.value == sixth_button) {
        delay(50); 
        digitalWrite(motorDirPin, HIGH);
        analogWrite(motorSpeedPin,75); 
        SerialMonitorInterface.println("Speed_6_opposite_direction");
     delay(50);
      }  

      else if (results.value == seventh_button) {
        delay(50); 
        digitalWrite(motorDirPin, HIGH);
        analogWrite(motorSpeedPin,0); 
        SerialMonitorInterface.println("Speed_7_highest_opposite_direction");
     delay(50);
      } 
      
      else {
        SerialMonitorInterface.print("Unrecognized code! ");
        SerialMonitorInterface.println(results.value, HEX);
      }
    } else {
      SerialMonitorInterface.print(results.decode_type);
      SerialMonitorInterface.print(" ");
      SerialMonitorInterface.print(results.bits);
      SerialMonitorInterface.print(" ");
      SerialMonitorInterface.println(results.value, HEX);
         }
  }
}
\end{lstlisting}

~\\


\end{document}